\documentclass[10pt,a4paper,twocolumn,english,prb,aps,showpacs,floatfix,groupedaddress,superscriptaddress,longbibliography]{revtex4-1}
\usepackage{graphicx}
\usepackage{epsfig}
\usepackage[english]{babel}
\usepackage{amsmath}
\usepackage{amssymb}
\usepackage{amsfonts}
\usepackage{longtable}
\usepackage{dcolumn}
\usepackage{bm}
\usepackage{bbm}
\usepackage{color,array}
\usepackage{colortbl}
\usepackage{dsfont}
\usepackage{mathtools}
\usepackage{subfigure}  
\usepackage[breaklinks,colorlinks=true,urlcolor=blue,citecolor=blue,linkcolor=blue]{hyperref}
\usepackage{breakurl}

\begin{document}

\newcommand{\be}{\begin{equation}}
\newcommand{\ee}{\end{equation}}
\newcommand{\bearr}{\begin{eqnarray}}
\newcommand{\eearr}{\end{eqnarray}}
\newcommand{\bseq}{\begin{subequations}}
\newcommand{\eseq}{\end{subequations}}
\newcommand{\nn}{\nonumber}
\newcommand{\dagg}{{\dagger}}
\newcommand{\vpdag}{{\vphantom{\dagger}}}
\newcommand{\bpm}{\begin{pmatrix}} 
\newcommand{\epm}{\end{pmatrix}} 
\newcommand{\bs}{\boldsymbol}

\title{Interaction-driven topological phase transitions in fermionic SU($3$) systems}

\author{Mohsen Hafez-Torbati}
\email{torbati@itp.uni-frankfurt.de}
\affiliation{Institut f{\"u}r Theoretische Physik, Goethe-Universit{\"a}t,
60438 Frankfurt/Main, Germany.}

\author{Jun-Hui Zheng}
\affiliation{Institut f{\"u}r Theoretische Physik, Goethe-Universit{\"a}t,
60438 Frankfurt/Main, Germany.}
\affiliation{Center for Quantum Spintronics, Department of Physics, Norwegian University of Science and Technology, 
NO-7491 Trondheim, Norway}

\author{Bernhard Irsigler}
\affiliation{Institut f{\"u}r Theoretische Physik, Goethe-Universit{\"a}t,
60438 Frankfurt/Main, Germany.}

\author{Walter Hofstetter}
\email{hofstett@physik.uni-frankfurt.de}
\affiliation{Institut f{\"u}r Theoretische Physik, Goethe-Universit{\"a}t,
60438 Frankfurt/Main, Germany.}

\date{\rm\today}

\begin{abstract}
We consider SU($3$) fermions on the triangular lattice in the presence 
of a gauge potential which stabilizes a quantum Hall insulator (QHI) at the 
density of one particle per lattice site. We investigate 
the effect of 
the Hubbard interaction, favoring magnetic long-range order, and a three-sublattice 
potential (TSP), favoring a normal insulator (NI), on the system. 
For weak TSP we find that the Hubbard interaction drives the QHI into 
a three-sublattice magnetic Mott insulator (MMI).
For intermediate values of TSP we identify two transition points upon increasing the 
Hubbard interaction. The first transition is from the NI to the QHI and the second
transition is from the QHI to the MMI. 
For large values of the TSP a charge-ordered magnetic insulator (COMI) 
emerges between the NI and the QHI, leading to an interaction-driven COMI-to-QHI 
transition. 
\end{abstract}


\maketitle

\section{introduction}
Since the experimental discovery of the quantum Hall effect in two-dimensional (2D) 
electron systems \cite{Klitzing1980} novel types of band insulators such as 
quantum Hall (QHI) \cite{Thouless1982} and quantum spin Hall (QSHI) 
insulator \cite{Kane2005}
have been identified, which are characterized by topological invariants and can not be 
adiabatically connected to the previously-known normal insulators (NIs) 
\cite{Hasan2010}. The QHI
occurs at particular particle fillings when a constant magnetic field  is applied perpendicular 
to a 2D lattice potential, splitting a single energy band into several 
subbands \cite{Hofstadter1976}, each one carrying an integer quantum number 
\cite{Thouless1982} called Chern number \cite{Simon1983}. The QSHI is a result 
of time-reversal symmetry and spin-orbit coupling
and is characterized by a $\mathds{Z}_2$ topological invariant \cite{Kane2005a}.

The effect of interaction on a band insulator (BI) and emergence of Mott physics 
in the strong coupling regime has been an interesting problem 
for a long time \cite{Nagaosa1986a}, initially motivated by the observation of 
neutral-ionic phase 
transition in organic compounds \cite{Torrance1981a}.
A spontaneously dimerized phase \cite{Fabrizio1999,Manmana2004,Batista2004,Loida2017}
stabilized by condensation of a singlet exciton 
\cite{Hafez2014,Hafez2015,Hafez2010b,Hafez2011} separates 
the NI from Mott insulator (MI) as is studied via the 1D ionic Hubbard model. 
The ground state phase diagram of the 2D model is controversial \cite{Paris2007,Kancharla2007,Hafez-Torbati2016}.

In recent years, there has been a large interest in interacting 
topological insulators \cite{Rachel2018}, with a focus on realizing 
topological many-body quantum states such as fractional QHI \cite{Parameswaran2013} 
and studying interaction-driven topological phase transitions 
\cite{Cocks2012,Budich2013,Amaricci2015,He2011,Vanhala2016,Jiang2018}. 
In the time-reversal-invariant Harper-Hofstadter-Hubbard model
with a 
spin-mixing hopping term
an 
interaction-driven NI-to-QSHI transition is identified \cite{Cocks2012}, which is found also 
in an extended Bernevig-Hughes-Zhang-Hubbard model \cite{Budich2013,Amaricci2015}.
The competition of the Hubbard interaction and the staggered potential in the Haldane-Hubbard model stabilizes 
an antiferromagnetic Chern insulator (AFCI) 
where one of the spin components is in the quantum Hall and the other in 
the normal state \cite{He2011,Vanhala2016}. Such an AFCI 
is proposed also for the Kane-Mele-Hubbard model but with a spontaneous breaking of 
the time-reversal symmetry \cite{Jiang2018}.

Spin-orbit coupling in multicomponent systems can give rise to a 
richer topological band structure compared to the SU($2$) case \cite{Barnett2012,Bornheimer2018,Yau2019}.
In the Mott regime SU($N$) systems 
are potential candidates to find  novel 
ordered and disordered MIs \cite{Gorshkov2010,Toth2010,Zhou2016,Nataf2016,Hafez-Torbati2018,Chung2019}. 
Furthermore, interaction-driven 
metallic phases and a charge-ordered magnetic insulator (COMI) are reported as 
a result of competing charge and magnetic order in fermionic SU($3$) systems \cite{Hafez-Torbati2019}.

Here we investigate SU($3$) fermions on the triangular lattice
at $1/3$ filling in the presence of a gauge potential stabilizing 
a QHI. We study the effect of the Hubbard interaction and a three-sublattice 
potential (TSP) on the QHI phase. For weak TSP, the Hubbard interaction drives
the QHI into a three-sublattice magnetic MI (MMI). 
For intermediate values of the TSP we find the NI at weak and the MMI at strong 
Hubbard $U$, separated by a QHI.
For large TSP an additional COMI phase emerges between the NI and the QHI. 
This leads to the realization of an interaction-driven COMI-to-QHI transition.
The study is experimentally motivated by the recent progress in realization 
of artificial gauge fields \cite{Aidelsburger2013,Miyake2013,Jotzu2014,Aidelsburger2018a} and creation of SU($N$)-symmetric 
multicomponent systems \cite{Ottenstein2008,Huckans2009,Hara2011,Taie2012,Cazalilla2014} 
in optical lattices.
The Hamiltonian reads
\bearr
\label{eq:hamiltonian}
H\!=&-&t\sum_{\left\langle {\bs r} {\bs r}' \right\rangle}\sum_{ \alpha} 
\left( e^{2\pi i \phi_{\bs{r},\bs{r}'}}
c^\dagg_{\bs{r}'\alpha} c^\vpdag_{\bs r \alpha} + {\rm H.c.}
\right)
+
\sum_{\bs r\alpha} \Delta_{\bs{r}}^\vpdag n^\vpdag_{\bs{r}\alpha}
\nn \\
&+&  
U\sum_{\bs r} \sum_{\alpha<\alpha'} \! 
n^\vpdag_{\bs{r}\alpha}  n^\vpdag_{\bs{r}\alpha'} 
\quad,
\eearr
where $c^\dagg_{{\bs r} \alpha}$ is the fermionic creation operator at the 
lattice position 
${\bs r}$ with the spin component $\alpha$, 
$n^\vpdag_{\bs{r}\alpha}\!=\!c^\dagg_{{\bs r} \alpha} c^\vpdag_{{\bs r} \alpha}$
is the occupation number operator,
and the summation over $\left\langle {\bs r} {\bs r}' \right\rangle$ 
restricts the hopping to nearest-neighbor sites. 
The hopping phase factors $\phi_{\bs{r},\bs{r}'}$  around each triangle 
add up to a constant $\Phi$ which describes the magnetic flux  
going through each triangle in units of the magnetic flux quantum. 
The three sublattices $A$, $B$, and $C$ of the 
tripartite triangular lattice acquire respectively the onsite energies 
$-\Delta_1$, $0$, and $+\Delta_2$ due to the second term, the TSP. 
The last term 
is the Hubbard interaction.

\section{technical aspects}

We map the triangular lattice to the square lattice with
hopping along the $\hat{\bs x}$, $\hat{\bs y}$, and $(\hat{\bs x}+\hat{\bs y})$ directions.
We consider the hopping phase factors $\phi_{{\bs r},{\bs r}+\hat{\bs x}}=0$, 
$\phi_{{\bs r},{\bs r}+\hat{\bs y}}=(2m+2n+1)\Phi$, and 
$\phi_{{\bs r},{\bs r}+\hat{\bs x}+\hat{\bs y}}=2(m+n+1)\Phi$ from the lattice 
position ${\bs r}=ma\hat{\bs x}+na\hat{\bs y}$, where $a$ is 
the lattice constant and $m,n \in \mathds{Z}$ \cite{phasefactor}. 
There are three sites in the unit cell for $\Phi=1/6$, which is the flux
we consider in this paper. 
In the absence of interaction the Hamiltonian reduces to a 
three-level problem in momentum space leading to three 
distinct Bloch bands with a three-fold spin degeneracy each. 
We determine the Chern number of the system at $U=0$ using twisted boundary 
conditions \cite{Niu1985,Kudo2019}.
We employ real-space dynamical mean-field theory (DMFT) 
\cite{Potthoff1999,Song2008,Snoek2008} which we 
implemented for SU($N$) systems in Ref. \onlinecite{Hafez-Torbati2018} 
to address the Hamiltonian at 
finite $U$. 
In real-space DMFT the self-energy is approximated to be local 
but it can be position-dependent. 
We consider $L\times L$ lattices with $L=30$ and periodic boundary 
conditions unless mentioned otherwise. 
We use the exact diagonalization (ED) impurity solver with four and five 
bath sites and check that the results nicely agree across different 
transition points. The presented results are for five bath sites unless 
mentioned otherwise.
We have used the inverse temperature $\beta=32/t$. 
We find at different selected 
parameter values
that the results remain unchanged compared to the ones obtained 
using a zero temperature ED impurity solver
\cite{Caffarel1994}.
We expect that a temperature $T=t/32$ is low enough to capture 
the ground state properties of the model.

We evaluate the Chern number of the interacting system 
using the topological Hamiltonian approach \cite{Wang2012}. 
This method states that  
the Chern number of an interacting system is equal to  
the Chern number of an effective non-interacting model called 
``topological Hamiltonian'', which in the Bloch form reads
\be
\label{eq:toph}
h_t({\bs k})=h_0({\bs k})+\Sigma({\bs k},i\omega=0),
\ee
where $h_0({\bs k})$ is the non-interacting part of the 
original model and $\Sigma({\bs k},i\omega)$ stands 
for the self-energy.
In DMFT the self-energy is local and we have no element in 
the Hamiltonian and in the self-energy linking different spin components. 
Consequently, the effect of self-energy in Eq. \eqref{eq:toph}
will be to renormalize the TSP to
\bseq
\label{eq:rdelta}
\begin{align}
 \tilde{\Delta}_{1,\alpha}^\vpdag&=\Delta_1+\left(\Sigma_{B,\alpha}^\vpdag(0)
 -\Sigma_{A,\alpha}^\vpdag(0)\right),\\
 \tilde{\Delta}_{2,\alpha}^\vpdag&=\Delta_2+\left(\Sigma_{C,\alpha}^\vpdag(0)
 -\Sigma_{B,\alpha}^\vpdag(0)\right),
\end{align}
\eseq
up to an irrelevant shift in the energy spectrum. We have used 
$\Sigma_{A,\alpha}(0)$ for the zero-frequency self-energy on 
sublattice $A$ with spin component $\alpha$ and similarly for 
sublattices $B$ and $C$ \cite{zeroself}. 
The effective TSP Eq. \eqref{eq:rdelta} in paramagnetic phases 
is spin-independent, while in magnetically ordered phases, i.e., 
in phases with broken SU($3$) symmetry, it depends on the spin.
This shows that different spin components can in principle occur 
in distinct topological regions.

\section{results}
\begin{figure}[t]
    \centering
     \includegraphics[width=0.75\linewidth,angle=-90]{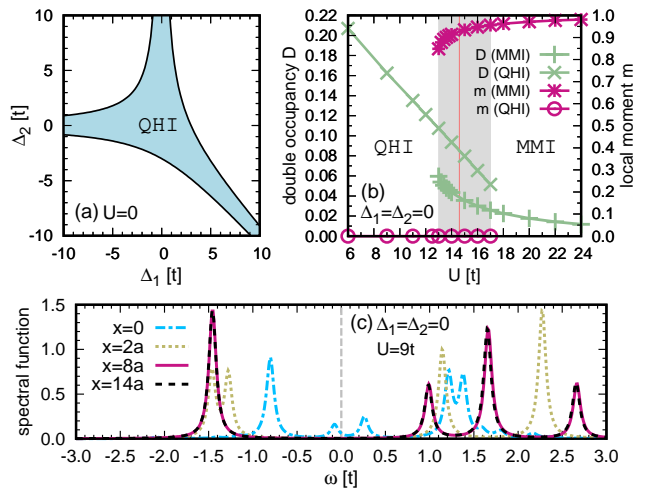}
     \caption{(a) 
     Phase diagram of the model for the Hubbard interaction $U\!=\!0$ in the $\Delta_1$-$\Delta_2$ plane. 
     The shaded area corresponds to the quantum Hall insulator (QHI) and 
     the white area to the normal insulator (NI). 
     (b) The double occupancy and the 
     local moment in the QHI and in the magnetic Mott insulator (MMI) versus $U$ at $\Delta_1\!=\!\Delta_2\!=\!0$. 
     The gray area is the coexistence region and the vertical solid line marks the transition point.
     (c) The spectral function $A_{{\bs r}\alpha}(\omega)$ at $U\!=\!9t$ and $\Delta_1\!=\!\Delta_2\!=\!0$ 
     versus frequency $\omega$ 
     for a cylindrical geometry with edges at $x=0$ and $x=29a$.
     }
     \label{fig:combined}
\end{figure}

Fig. \ref{fig:combined}(a) shows the phase diagram of the model at $U=0$ 
in the $\Delta_1$-$\Delta_2$ plane. The shaded area denotes the 
QHI and the white area the NI phase. 
In the QHI each spin component $\alpha$ contributes a Chern number $\mathcal{C}_{\alpha}=1$, 
leading to the Chern number $\mathcal{C}=3$ for the full system.
The three asymptotic branches for the phase boundaries can be understood based on the 
sublattice degeneracy. For instance,
sublattices $A$ and $B$  are degenerate at $\Delta_1=0$ and upon 
increasing $\Delta_2 \to +\infty$ always the two lowest Bloch bands remain topological, 
leading to a QHI state at $1/3$ filling.
Fig. \ref{fig:combined}(a) can be used to 
determine also the topological properties of the interacting model 
as the effect of the interaction is only to renormalize the TSP.

\begin{figure}[t]
    \centering
     \includegraphics[width=0.51\linewidth,angle=-90]{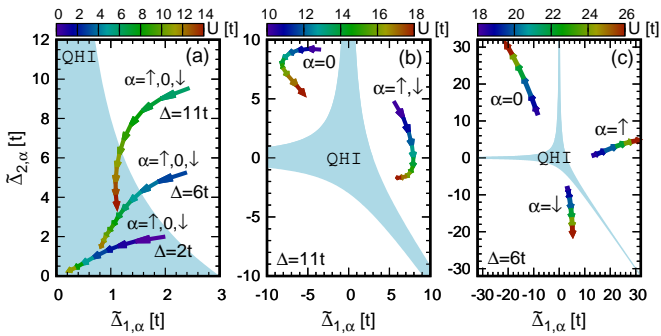}
     \caption{The evolution of the effective potential 
     as a function of $U$ for the paramagnetic DMFT solution (a) for the charge-ordered 
     magnetic insulator with $\Delta\!=\!11t$ (b) and for the magnetic Mott 
     insulator with $\Delta\!=\!6t$ (c) for the spin components $\alpha=\uparrow$, $0$, 
     and $\downarrow$.}
     \label{fig:D1D2}
\end{figure}

For SU($3$) systems 
we define the double occupancy 
$D_{\bs r}\!=\!\sum_{\alpha<\alpha'} \langle n_{{\bs r} \alpha}
 n_{{\bs r} \alpha'} \rangle$ 
and
the local moment 
$m_{\bs r}\!=\!
\sqrt{3}
 \left| \langle \mathcal{S}_{\bs r}\rangle \right|/2$  
where 
$\mathcal{S}_{\bs r}^{i}\!=\!\sum_{\alpha\alpha'}
c^\dagg_{\bs{r}\alpha} \lambda^{i}_{\alpha\alpha'} c^\vpdag_{\bs r \alpha'}$
for $i\!=\!1,\cdots,8$ define the elements 
of the eight-dimensional pseudospin operator $\mathcal{S}_{\bs r}$ 
with $\lambda^{i}$ being the Gell-Mann matrices. 
In magnetic phases there is a continuous degeneracy and 
we focus on the solution with pseudospin order in 
the $\hat{\mathcal{S}}_3$-$\hat{\mathcal{S}}_{8}$ plane.
In Fig. \ref{fig:combined}(b) the double occupancy and the local moment 
in the QHI and in the MMI are depicted versus $U$ for 
$\Delta_{1}\!=\!\Delta_{2}\!=\!0$.
The QHI and the MMI are two DMFT solutions coexisting in the 
gray area.
The QHI results from the zero effective TSP in the paramagnetic region.
The MMI is topologically trivial as we find all the three spin components in the normal state. 
This is a point which we will discuss further in Fig. \ref{fig:D1D2}.
For $\Delta_{1}\!=\!\Delta_{2}\!=\!0$ $D_{\bs r}$ and $m_{\bs r}$ are 
position-independent.
The red solid line 
at $U_c \simeq 14.5t$ 
specifies the 
transition point obtained by comparing the energy of the two states.
The MMI has a three-sublattice magnetic order such that on 
each sublattice one of the spin components has the dominant density 
and the density of the other two components is equal, 
leading to a $120^\circ$ pseudospin order \cite{Hafez-Torbati2018,Bauer2012}. 

To investigate gapless edge states in the QHI we consider a $30\times30$ lattice 
with periodic boundary condition along $\hat{\bs y}$ and open boundary condition along $\hat{\bs x}$, 
i.e., a cylindrical geometry, 
with edges at $x\!=\!0$ and $x\!=\!29a$. 
The spectral function at position ${\bs r}$ for the spin component $\alpha$ is defined 
from the local Green's function as 
$A_{{\bs r}\alpha}(\omega)\!=\!-\frac{1}{\pi}{\rm Im}G_{{\bs r}\alpha}(\omega+i\epsilon)$
where $\epsilon$ is a numerical broadening factor.
In Fig. \ref{fig:combined}(c)
the spectral function 
$A_{{\bs r}\alpha}(\omega)$
for $U\!=\!9t$ and $\Delta_{1}\!=\!\Delta_{2}\!=\!0$ is plotted versus frequency $\omega$ in the range $-3t\!<\!\omega\!<\!3t$
with $\epsilon\!=\!0.05t$. 
The dashed line at $\omega\!=\!0$ specifies the Fermi energy.
Due to the finite number of bath sites $N_{\rm b}\!=\!5$ in the impurity problem the fine details 
of the spectral function can not be reserved. However, one can clearly identify the spectral
contribution  from the edge $x\!=\!0$ near 
the Fermi energy, which vanishes upon approaching the bulk $x\!=\!14a$. 
It is interesting that even with a finite number of bath sites 
one can see evidence of gapless edge states. 
The edge and the bulk spectral function on finite clusters in an interacting topological insulator  
is discussed also in Ref. \cite{Varney2010}.
However, we notice that 
computing topological invariants is a more accurate and reliable way to recognize topological phase transitions.

\begin{figure}[t]
    \centering
     \includegraphics[width=0.74\linewidth,angle=-90]{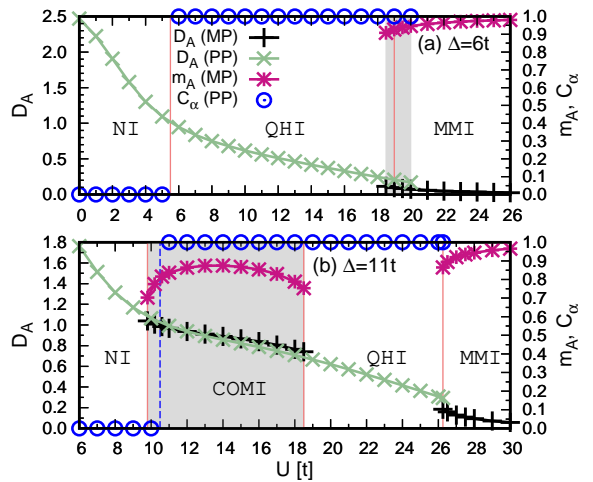}
     \caption{The double occupancy $D_{A}$ and the 
     local moment $m_{A}$ on sublattice $A$, and the Chern number $\mathcal{C}_{\alpha}$ 
     for the spin component $\alpha$ versus the Hubbard interaction $U$ for $\Delta\!=\!6t$ (a) 
     and $\Delta\!=\!11t$ (b). 
     The gray area denotes coexistence of magnetic and paramagnetic solutions.
     The local moment $m_{A}$ is given only in 
     magnetic phases (MP), i.e., in the charge-ordered magnetic insulator (COMI) and 
     in the magnetic Mott insulator (MMI). 
     The given spin-independent Chern number $\mathcal{C}_{\alpha}$ is 
     for paramagnetic phases (PP), i.e., for 
     the normal insulator (NI) and for the quantum Hall insulator (QHI), as it is zero for MP.
     The red solid lines mark the 
     transition points and the dashed blue line denotes the NI-to-QHI 
     transition ignoring the magnetic DMFT solution.}
     \label{fig:finiteD}
\end{figure}

We leave a general study of the Hubbard interaction on the phase diagram Fig. \ref{fig:combined}(a)
for future research and consider here for simplicity $\Delta_1\!=\!\Delta_2\!=:\!\Delta\!>\!0$. We believe that small deviations 
from this symmetric case will not change the physics discussed in the following essentially. 
At $U\!=\!0$ there is a transition from 
the QHI to the NI at $\Delta_c \!=\! 3t/\sqrt{2}$ upon increasing $\Delta$.
In Fig. \ref{fig:finiteD} we have plotted 
the double occupancy $D_{A}$ and the local moment $m_{A}$ on sublattice $A$ as well 
as the Chern number $\mathcal{C}_{\alpha}$ 
versus the Hubbard interaction $U$ for $\Delta\!=\!6t$ (a) and $\Delta\!=\!11t$ (b). 
To avoid a busy figure 
the local moment is given only in magnetic phases (MP) 
as it is trivially zero in paramagnetic phases (PP). 
In addition we find 
$\mathcal{C}_{\alpha}\!=\!0$ for all the three spin components 
in MP, see also below. 
The given spin-independent
$\mathcal{C}_{\alpha}$ is for PP.
The gray area indicates coexistence of magnetic and paramagnetic DMFT solutions. 
One notices that in Fig. \ref{fig:finiteD}(b) the COMI always coexists 
with a paramagnetic phase and the given Chern number is for the paramagnetic phase not for the COMI.
The red vertical solid line specifies the transition point and is obtained 
by comparing the energies of the two states in the case of coexistence. 
The blue vertical dashed line denotes the  NI-to-QHI transition 
in the case that paramagnetic solution is enforced. 

One can see from Fig. \ref{fig:finiteD}(a) that the Hubbard interaction 
drives the NI into the QHI and subsequently
the QHI into the MMI. Similar sequences of phase 
transitions are found in SU($2$) topological systems 
\cite{Cocks2012,Budich2013,Amaricci2015,He2011,Vanhala2016,Jiang2018}.
Upon increasing the TSP 
to $\Delta\!=\!11t$ in Fig. \ref{fig:finiteD}(b) 
a COMI phase emerges 
between the NI and the QHI. In the COMI phase, sublattice $A$ is almost doubly 
occupied with two spin components, sublattice $B$ is mainly occupied with the third component, 
and sublattice $C$ is almost empty. The local moment on sublattice $A$ and $B$ 
is equal and it is zero on sublattice $C$. 
There is a $180^\circ$ pseudospin order on sublattices $A$ and $B$ \cite{Hafez-Torbati2019}.
We find that the COMI always has a lower energy than the paramagnetic phases, i.e., 
the NI and the QHI are metastable.
We notice that charge order is 
an intrinsic property of the COMI phase as it is not adiabatically connected 
to any phase with a uniform charge distribution. This is to be compared with 
the QHI and MMI phases which are adiabatically connected to $\Delta\!=\!0$ limit 
where the charge distribution is uniform. 
We believe the Hubbard interaction driving a magnetic phase into 
a quantum Hall state as it occurs in the COMI-to-QHI transition is 
a peculiar feature of multicomponent systems which has no SU($2$) counterpart.

The double occupancy $D_{\!A}$ versus $U$ in Fig. \ref{fig:finiteD} exhibits 
a change of slope in different phases and 
can be conveniently measured in optical lattices using the photoassociation 
technique \cite{Taie2012}. 
The magnetic order can be identified using a quantum gas microscope 
\cite{Mazurenko2017, Brown2017}. 
Lower temperatures are accessible in multicomponent systems compared to 
the SU($2$) case due to a Pomeranchuk cooling effect \cite{Ozawa2018}. 
We notice that to realize magnetic order at finite temperature in our system a weak coupling 
in the third direction or an interaction anisotropy is required.

To further clarify the topological nature of different phases we study 
in Fig. \ref{fig:D1D2} the 
evolution of the effective TSP 
as a function of $U$ for the paramagnetic DMFT solution (a), for the COMI with 
$\Delta\!=\!11t$ (b), and for the MMI with $\Delta\!=\!6t$ (c). The direction 
of the curves 
are upon increasing $U$. 
The shaded area corresponds to QHI and the white area 
to NI. One sees from Fig. \ref{fig:D1D2}(a) that for $\Delta\!=\!2t$ the system 
is always in the QHI region but for $\Delta\!=\!6t$ and $\Delta\!=\!11t$ a NI-to-QHI 
transition occurs. Figs. \ref{fig:D1D2}(b) and \ref{fig:D1D2}(c) 
demonstrate that the COMI and the MMI are topologically trivial as all the three spin components 
$\alpha=\uparrow$, $0$, and $\downarrow$ are in the NI region. 
The larger the local moment is in the MMI and in the COMI in Figs. \ref{fig:finiteD}(a) 
and \ref{fig:finiteD}(b)
the deeper the corresponding topological Hamiltonian is in the  NI  in 
Fig. \ref{fig:D1D2}.
The interaction-driven topological phase transitions can be studied in 
optical lattices using the tomography scheme 
proposed in Ref. \onlinecite{Zheng2020}. 

\begin{figure}[t]
    \centering
     \includegraphics[width=0.8\linewidth,angle=0]{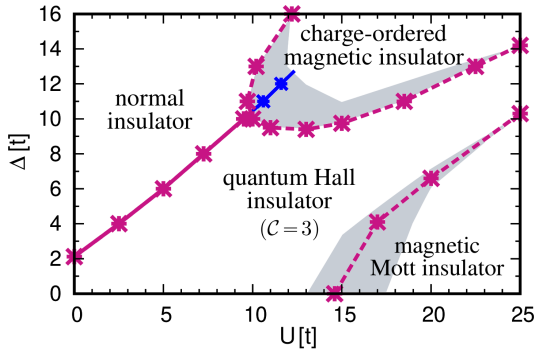}
     \caption{The phase diagram in the $U$-$\Delta$ plane. 
     The red lines denote the phase boundaries, 
     the gray areas represent the coexistence regions, and the blue line 
     separates the normal from the quantum Hall insulator when ignoring the 
     magnetic DMFT solution. The solid (dashed) line indicates a continuous 
     (discontinuous) transition.}
     \label{fig:PD}
\end{figure}

Fig. \ref{fig:PD} displays the phase diagram 
in the $U$-$\Delta$ plane. 
The gray areas denote the coexistence of magnetic and 
paramagnetic states, the red lines are the phase boundaries, 
and the blue line separates the NI from the QHI ignoring the magnetic DMFT solution.
The solid (dashed) line indicates a continuous (discontinuous) transition.
We have used four bath sites in the impurity problem due to the large number of 
data we needed to produce. However, by comparing Fig. \ref{fig:PD} with 
Fig. \ref{fig:combined}(b) and Fig. \ref{fig:finiteD} one can see the nice agreement 
for coexistence regions and transition points obtained with five and four bath sites.
We have performed further checks across some other selective transition points.
We always find that the NI-to-QHI transition is continuous, although discontinuous 
transitions in two-orbital systems are also reported \cite{Amaricci2015}. 
The coexistence regions shrink upon increasing $\Delta$. 
The QHI in the limit $U,\Delta\! \gg \!t$ appears around $U\!=\!2\Delta$ 
where the COMI and the MMI are degenerate in the atomic limit, 
i.e., at $t\!=\!0$ \cite{Hafez-Torbati2019}. 
We have produced the phase diagram up to $U\!=\!32t$ and $\Delta\!=\!20t$ and the QHI 
persists with a constant width. This width is proportional to $t$ and vanishes 
in the atomic limit.

\section{Summary and outlook}

To summarize, in recent years there has been a large interest in fermionic SU($N$) 
systems \cite{Gorshkov2010,Cazalilla2014} 
as well as in artificial gauge fields \cite{Aidelsburger2018,Cooper2019,Hofstetter2018} 
due to their possible realization in 
optical lattices. While studies of SU($N$) systems have mainly been focused on topological states in the absence 
of interaction \cite{Barnett2012,Bornheimer2018,Yau2019} and on Mott states in the strong 
coupling limit \cite{Gorshkov2010,Toth2010,Nataf2016,Zhou2016,Hafez-Torbati2018,Chung2019}, less attention has so far been 
paid to the competition of band and Mott insulator and possible 
emergence of 
intermediate phases and novel phenomena. 
This requires tuning the interaction from weak to strong 
which can experimentally be achieved by Feshbach resonances \cite{Inouye1998,Courteille1998,Bloch2008}.
In this paper we show that local correlations, which 
  are best known for the famous Mott transition, can drive a magnetic 
  phase into a quantum Hall state in multicomponent systems. $\mathds{Z}_2$ 
  lattice gauge theories are recently simulated using ultracold atoms in optical 
  lattices \cite{Barbiero2019,Schweizer2019}. Our work sets the stage for a generalization of
  static gauge fields with interactions to the dynamical case and for studies of
  $\mathds{Z}_3$ lattice gauge theories, which are linked to important issues 
  in high-energy physics.

\section{acknowledgement}
We would like to thank J. Panas for useful discussions. 
This work  was  supported  by  the  Deutsche  Forschungsgemeinschaft 
(DFG,  German  Research  Foundation)  under  Project No.  277974659  
via  Research  Unit  FOR  2414. This  work was  also  supported  by  
the  DFG  via  the  high  performance computing center LOEWE-CSC.


\begin{thebibliography}{70}
\bibitem [{pha()}]{phasefactor}%
  \BibitemOpen
  \href@noop {} {\bibinfo  {journal} {These phase factors can be obtained using
  the Peierls substitution for the gauge field ${\bs A}=B(x+y)\hat{\bs y}$
  which describes the magnetic field ${\bs B}=B\hat{\bs z}$. The advantage of
  the chosen gauge over the Landau gauge is that it makes the hopping phase
  factors depend on $(x+y)$, which is the same position dependency as in the
  three-sublattice potential, and consequently reduces the number of sites in
  the unit cell}\ }\BibitemShut {NoStop}%
\bibitem [{zer()}]{zeroself}%
  \BibitemOpen
  \href@noop {} {\bibinfo  {journal} {We find that the real-part of the
  self-energy at the smallest (in absolute value) Matsubara frequency
  accurately describes the zero-frequency self-energy obtained by a polynomial
  fit}\ }\BibitemShut {NoStop}%
\end{thebibliography}
\end{document}